\documentclass[onecolumn,epsfig,pre]{revtex4}
\usepackage{graphicx}
\usepackage{amssymb}
\usepackage{amsmath}
\usepackage{hyperref}
\usepackage{latexsym}

\def\be{\begin{equation}}
\def\ee{\end{equation}}
\def\bea{\begin{eqnarray}}
\def\eea{\end{eqnarray}}

\begin{document}

\title{Stabilization of overlapping biofilaments by passive crosslinkers}
%\shorttitle{Motor driven transport}
\author{Sougata Guha$^{1,2}$, Subhadip Ghosh$^{2}$ Ignacio Pagonabarraga$^{3}$ and Sudipto Muhuri$^{1}$}

\affiliation{$^{1}$Department of Physics, Savitribai Phule Pune University, Ganeshkhind, Pune 411007, India\\
$^{2}$Indian Institute of Science Education and Research Mohali, Knowledge City, Sector 81, SAS Nagar - 140306, Punjab, India.\\
$^{3}$CECAM, Centre Europ\'een de Calcul Atomique et Mol\'eculaire, \'Ecole Polytechnique F\'ed\'erale de Lasuanne, Batochime, Avenue Forel 2, 1015 Lausanne, Switzerland}

\begin{abstract}
The formation, maintenance and reorganization of the cytoskeletal filament network is essential for a number of  cellular processes. While the crucial role played by  active forces generated by motor proteins has been  studied extensively, only recently the importance of  passive forces exerted by non-enzymatic crosslinkers  has been realized. The interplay between active and passive proteins  manifests itself, e.g.,   during cell division, where the spindle structure formed by overlapping microtubules  is subject to both active sliding forces generated by crosslinking motor proteins and passive forces exerted by  passive crosslinkers, such as Ase1 and PRC1. We propose a minimal model to describe the stability behaviour of a pair of anti-parallel overlapping microtubules  resulting from the competition between active motors and  passive crosslinkers. We obtain the  stability diagram which characterizes the formation of stable overlap of the MT pair, identify the controlling biological parameters which determine their stability, and study the impact of mutual interactions between  motors and  passive crosslinkers on the stability of these overlapping filaments.  
\end{abstract}

\maketitle
The structural integrity of an eukaryotic cell is provided by the cytoskeleton filament network comprised of microtubules (MTs) and actin filaments \cite{cell}. The ability of the cell to reorganize itself to develop spatio-temporal   structures out of these cytoskeletal filaments is vital for the process of cell division, morphogenesis and movement \cite {cell,howard}. This ability hinges crucially on the activity of the motor proteins which attach and move on  these cellular filaments and generate active stresses that cause reorganization of the cytoskeletal filaments \cite{howard, nedelec, menon, madan, raja1, sm1,sm2, karsten1, debasoft}. A paradigmatic example in this context is the formation of a spindle  during mitotic cell division. The spindle structure comprises of overlapping anti-parallel MTs originating from centrosome poles within the cell. These overlapping arrays of anti-parallel MTs are subject to active sliding forces exerted by bipolar motor proteins such  as Eg5 and  kinesin-8 \cite{raja1}  which attach simultaneously to the two overlapping filaments and slide them with respect to each other. These molecular motors use the stored chemical energy from ATP hydrolysis to generate the sliding forces,  driving the cytoskeleton out of equilibrium. While {\it in-vitro} and {\it in-vivo} studies of mitotic spindle have established the centrality of the role played by the molecular motors in the organization of the cytoskeleton network \cite{kinesin1, kinesin2, karsten1, sm1,sm2, debasoft}, relatively less attention has been paid to the role of passive filament crosslinkers such as Ase1, PRC1 and MAP65 \cite{lansky1, dogic_review,lk1,lk2,lk3}. While it has been known earlier that these passive crosslinkers provide structural integrity of the filament network and  increases the effective friction between the sliding filaments, recent experiments have also shown that these entities are capable of generating entropic forces by harnessing the thermal energy from the cellular environment and are also involved in determining the stability and reorganization of the cytoskeletal filament network along with the molecular motors \cite{lansky1,lansky2, lansky3, dogic_review}.

Two mechanisms that lead to entropic forces between filaments have been identified in out of equilibrium cytoskeletal filament networks. On one hand, in a suspension of cytoskeletal filaments and small non-adsorbing polymers, such as Polyethylene Glycol (PEG), the volume occupied by the filaments is unavailable for the depletant PEG polymers. Decreasing the excluded volume due to the filaments which is achieved by filament bundling  and maximizing the overlap of the filaments, is entropically favorable. The gain in free energy for two overlapping filaments is proportional to their overlap length, $l$, and the associated entropic force acting along the long axis of the two filaments is constant, independent of the $l$,  set by the osmotic pressure  of the depletant polymers and associated geometric factors~\cite{dogic_prl,lansky3}.  On the other hand, overlapping microtubules in a mitotic spindle confine  freely diffusing, passive crosslinkers, such as Ase1 and MAP65~\cite {dogic_review, lansky3}. As a result, the confined proteins exert an effective pressure which  tends to increase $l$, with a strength that increases linearly in the number of crosslinkers and inversely proportional to $l$~\cite{lansky3}. These studies have highlighted the relevance of entropic forces in biological networks~\cite{dogic_review}, which are relevant also to understand the  self-assembly of biofilament bundles~\cite{dogic_prl}, and shown to be potentially relevant in spindle geometries~\cite{kruse_passive, karsten2}.

In this letter we identify an alternative  mechanism for the generation of  sliding forces by passive crosslinking proteins.  In contrast to previous mechanisms~\cite{lansky1, dogic_review, lansky3}, we take into account the exchange of passive protein crosslinkers with the environment in which the  filaments are suspended.  As a result, if  the evolution of the overlap length, $l$, is much slower than the exchange rate of the passive crosslinkers with the environment, the passive crosslinkers decrease the effective free energy of the complex by an amount that increases linearly with $l$. As we will show, passive crosslinkers induce an entropic force that increases with their density, $\phi$, independent of $l$, unlike the entropic forces exerted by confined passive crosslinkers~\cite{lansky1, dogic_review, debasoft}.

We first describe the model and obtain the  phase diagram associated with the linearly stable behaviour of a pair of overlapping anti-parallel MTs. We then identify the relevant biological parameters which control the system stability.  We also consider the stability of overlapping MTs for the situation where the  passive crosslinking proteins in the overlap interact with the active motor crosslinkers. 
%We contrast our findings with that of the situation where  passive crosslinkers  exchange with the medium is prevented \cite{lansky1,debasoft}. 

\section{Model}
\begin{figure}[t]
\begin{center}
\includegraphics[scale = 0.3]{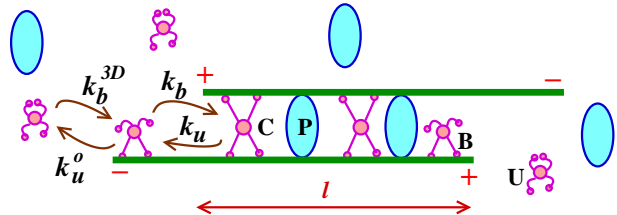}
\caption{(color online) 
Schematic representation of the system comprising of two overlapping anti-parallel microtubules with passive crosslinkers (P), motors which is bound to only one filament (B), crosslinked motors (C)  which are bound to both the filaments, and the unbound motors (U). The motors transform from bound state to crosslinked state with rate $k_b$ and from crosslinked state to bound state with rate $k_u$. $k_{u}^{o}$  is the unbinding  rate of the motor from the MT filament and $k_b^{3D}$ is the binding rate of free motors to the filament.}
\label{fig:1MT}
\end{center}
\end{figure}

We consider a pair of anti-parallel microtubules with an overlap length $l$ as depicted in  Fig.1. In the overlap region, kinesin motors can crosslink anti-parallel MTs and  slide them away from each other, generating an active outward force which tends to decrease the overlap length $l$. The  motor proteins (un)binding kinetics in the overlap region is expressed in terms of  rate equations for two populations of motors, e.g; $n_c$ which is the average number of motors crosslinked to the MT filaments, and $n_b$ which is the average number of motors when they are attached to only one of the two overlapping filament. The passive crosslinkers have a  linear density, $\phi$, in the overlapping region of the two filaments, set by the Langmuir kinetic process of attachment and detachment onto the filaments.  First we describe the case corresponding to the situation where the passive crosslinkers do not interact with the motor proteins. Subsequently we look at the situation where these passive crosslinkers also interact with motor proteins.

\subsection{Non-interacting motor -passive crosslinkers - MTs:} The binding  free energy of the passive crosslinkers to the microtubule, $-\epsilon_{o}$ , induces a net inward force on the pair of anti-parallel MT, $F_{p}\equiv -\frac{\partial E}{\partial l}$, associated with the total free energy reduction, ${\it E} = -\epsilon_{o}\phi l$, due to the passive crosslinkers.
We consider the relative motion of a pair of anti-parallel filaments. While in general  under {\it in-vivo} conditions the $(-)$ end of the filament would be coupled to the cell cortex, we focus our attention only on the overlap dynamics resulting from the interplay of the motor proteins and the forces exerted by the passive crosslinkers in the overlap region. In the absence of the crosslinkers, the overlapping filaments will slide with respect to each other with a velocity $2v_{o}$, where $v_o$ is the motor velocity of the crosslinking motors in the absence of any external load. The load force exerted by the passive crosslinkers on the molecular motors will reduce their average sliding velocity. Assuming a linear force-velocity relation for the motor velocity with equal load sharing between the crosslinked motors, the overlap length  $l$ evolves as,
\begin{equation}
\frac{dl}{dt} = -2v_{o}\left(1 - \frac{F_p}{n_{c}f_{s}}\right) +\frac{1}{\Gamma} F_{p}
\end{equation}
where,  $f_s$ stands for the stall force for a single motor and $\Gamma$ is the friction coefficient associated to the motion of the filaments against their embedding medium~\cite{karsten2}.  

Due to the passive force experienced by the crosslinked motors, the unbinding rate of the motors from the crosslinked state to single filament bound state  grows exponentially with the  load force felt by the individual motors \cite{ignaref}. The corresponding dynamics of the average number crosslinked motors $n_c$ can then be expressed as, 
\begin{equation}
\frac{dn_c}{dt} = k_{b}n_{b} - k_{u}^{o}n_{c}\exp\left[\frac{b F_{p}}{n_{c} k_{B}T}\right]
\end{equation}
where, $k_{b}$ and $k_{u}^{o}$ are the rates with which the uncrosslinked motors are converted to crosslinked state and vice-versa in the absence of external load force, and $b$ is the  length scale characterizing the activation process leading to unbinding from the crosslinked state of the motor. 

The number of bound motors in overlapped region which are not  crosslinked, $n_b$,   evolves due to interconversion between bound  and crosslinked motor states,  (un)binding process of free unbound motor on to the filament, and  the incoming flux of bound motors $2J$ from the two boundaries of the overlap region. When the overlap length is much smaller than the MT length, $J =  \frac{ k_{b}^{3d}\rho_{3d}}{k_{u}^{o}}\left[\frac{v_{o}F_{p}}{n_{c}f_{s}} + \frac{F_p}{\Gamma}\right]$ \cite{ignaref, sm1}. Thus the dynamic equation for $n_{b}$ can be expressed as, 
\begin{eqnarray}
\frac{dn_b}{dt} &=& k_{u}^{o}n_{c}\exp\left[\frac{b F_{p}}{n_{c}k_{B}T}\right] - (k_{u}^{o} + k_{b})n_{b} + k_{b}^{3D}\rho_{3d} l \nonumber\\
&+& \frac{2 k_{b}^{3d}\rho_{3d}}{k_{u}^{o}}\left[\frac{v_{o}F_{p}}{n_{c}f_{s}} + \frac{F_p}{\Gamma}\right]
\end{eqnarray}
where, $\rho_{3d}$ is the density of suspended motors in the bath, and $k_{b}^{3d}$ is the binding  rate of the free motors to the filament in the overlap region.

The  behaviour of a pair of anti-parallel filaments that is subject to sliding forces due to motors and passive crosslinkers is determined by $n_c$, $n_b$ and $l$ , and correspondingly, Eqs.(1)-(3) describe the dynamics of the composite composed of filaments, motors and passive crosslinkers. We rescale time in units of $1/k_{u}^{o}$, overlap length in terms of the average run length of single motor,  $l_{p} = v_{o}/k_{u}^{o}$ and forces in terms of  $k_BT/b$. We also introduce a  length scale associated with the  characteristic forces experienced by the passive crosslinkers, $l_{e} = \epsilon_{o}b/k_{B}T$. We define dimensionless variables $\tau=t k_u^o$, $\tilde{f_s} = b f_s/k_{B}T$, $\tilde l = \tilde f_{s} l/l_p$, $\tilde \Gamma = \frac{2v_{o}\Gamma b}{k_{B} T}$ and $\phi_o = l_e\phi$. Then Eqs. (1)-(3) can be re-expressed in terms of these dimensionless variables as,  
\begin{eqnarray}
\frac{d\tilde l}{d\tau} &=& \frac{2 \phi_o}{n_c} + 2\tilde f_{s}\left(\frac{\phi_o}{\tilde \Gamma} - 1 \right)\\
\frac{d n_c}{d \tau} &=& \gamma n_{b} - n_{c}\exp\left(\frac{\phi_o}{n_{c}}\right) \\  
\frac{d n_b}{d \tau}&=&n_{c}\exp\left(\frac{\phi_o}{n_{c}}\right)  + \Delta_n\left( \tilde l  + \frac{2\phi_o}{n_{c}} + \frac{4\tilde f_s \phi_o}{\tilde \Gamma}\right)\nonumber\\
&-& (1 + \gamma) n_{b}
\end{eqnarray}
We identify the relevant control parameters associated with motor-filament interaction. The linear density of the passive proteins, $\phi_o$,  is one of the biologically relevant control parameters. We also identify $\Delta_n = \frac{\rho_{3d}k_{b}^{3d}v_o}{\tilde f_{s} k_{u}^{o}k_{u}^{o}}$, which measures the strength of motor flux reaching the overlap region, and $\gamma = \frac{k_b}{k_{u}^{o}}$, which quantifies the  asymmetry in motor attachment/detachment  at vanishing load force, as relevant biological control parameters.

\subsection{Interacting motor -passive crosslinkers - MTs:} 

Until now, in formulating the evolution equations governing the dynamics of overlapping filaments, we have not taken into account the mutual interaction between the passive crosslinkers with the motors. However in general  the passive crosslinker proteins will interact with  motor proteins \cite{lk1,lk2,lk3,kruse_passive}. For instance, in experiments with reconstituted system comprising of MT and crosslinking protein PRC1, it has been observed that  these crosslinking proteins had an effective attractive interaction with kinesin motors in the overlap region \cite{lk2,lk3}. We quantify the interaction strength  between   passive proteins and motor crosslinkers with a coupling parameter, $a$, in terms of which we propose a simple interaction free energy ${\it E} = -\phi l (\epsilon_{o}+ a n_{c} + a n_{b})$ and a corresponding  force, $F_{p}^{*} = \phi(\epsilon_{o}  + a n_{c} + a n_{b})$.  Although it  has been argued on theoretical grounds that excluded volume interactions between diffusing passive proteins and motors can lead to phase segregation in the overlap region \cite{kruse_passive}, for the sake of simplicity,  we consider homogeneous admixture of motors and passive crosslinkers in the overlap region and focus on the impact of the simplest coupling between the passive crosslinkers and the  molecular motors. We quantify this coupling through the dimensionless coupling constant $\alpha_{o} \equiv a b /l_e  k_{B} T$. Accordingly, the relevant dimensionless force becomes  $\tilde F_{p}^{*} =  \phi_{o} + \alpha_o \phi_o  n_c + \alpha_o \phi_o  n_b$. While positive values of $\alpha_o$ correspond to a repulsive interaction between the passive crosslinkers and the motors, negative values of $\alpha_o$ corresponds to an attractive interaction between the crosslinking motors and the passive proteins.

The corresponding scaled equations of motion describing the dynamics of the system read as
\begin{eqnarray}
\frac{d\tilde l}{d\tau} &=& \frac{2 \phi_o}{n_c} + 2\tilde f_{s}\left(\frac{\phi_o}{\tilde \Gamma} - 1 \right)\\ &+& 2\alpha_{o}\phi_{o}\left( 1 + \frac{n_b}{n_c} + \frac{\tilde f_{s} (n_c + n_b)}{\tilde \Gamma} \right)\nonumber \\
\frac{d n_c}{d \tau} &=& \gamma n_{b} - n_{c}\exp\left(\frac{\phi_o}{n_{c}} + \alpha_{o} \phi_{o} + \alpha_{o}\phi_{o}\frac{n_b}{n_c} \right)\\  
\frac{d n_b}{d \tau} &=& -(1 + \gamma) n_{b} + n_{c}\exp\left(\frac{\phi_o}{n_{c}} + \alpha_{o} \phi_{o} + \alpha_{o}\phi_{o}\frac{n_b}{n_c} \right) \nonumber\\
&+& \Delta_n\left( \tilde l  + \frac{2\phi_o}{n_{c}} + \frac{4\tilde f_s \phi_o}{\tilde \Gamma}\right) \nonumber\\&+& 2\Delta_{n}\alpha_{o}\phi_{o}\left( 1 + \frac{n_b}{n_c} + \frac{2\tilde f_{s}}{\tilde \Gamma}(n_c + n_b)  \right).
\end{eqnarray}
Equations (7)-(9) describe the dynamical behavior of the system when the  motors interact with  passive proteins. In order to analyze the stability behavior for this system ( with and without mutual interaction between crosslinking motors and passive proteins), we first obtain steady state solutions for $n_c$, $n_b$ and $\tilde l$ and perform a linear stability analysis about the fixed point to obtain the stability diagram for obtaining a stable overlap of the anti-parallel MTs. For the case of motors interacting with the passive crosslinkers, the fixed points can be obtained numerically and the corresponding stability curve separating regions of stable overlap with the unstable region is also obtained numerically. If the passive proteins interact preferentially with the crosslinked motors alone then the corresponding sliding force acting on the overlapping filaments simplifies to $F_{p}^{*}  = \epsilon_{o}\phi + a\phi n_{c}$.  Eqs.(7)-(9) simplify accordingly, and  analytical expression for the the fixed points and the corresponding linear stability boundaries can be obtained.

\section{Results}
We analyze the stability of a   complex  composed by   overlapping anti-parallel MT  that traps kinesin motors and passive crosslinkers.  We analyze the impact that the different  types of interactions induced by the passive cross-linker have on the stability of the MT-motor complex as a function of the relevant biological control parameters. Eqs.(7)-(9)  highlights the  relevance of $\Delta_n$ and concentration of passive crosslinkers $\phi_o$, and we focus on their impact  on the mechanical properties of these active complexes.
\begin{figure}[t]
\begin{center}
\includegraphics[scale = 1.0]{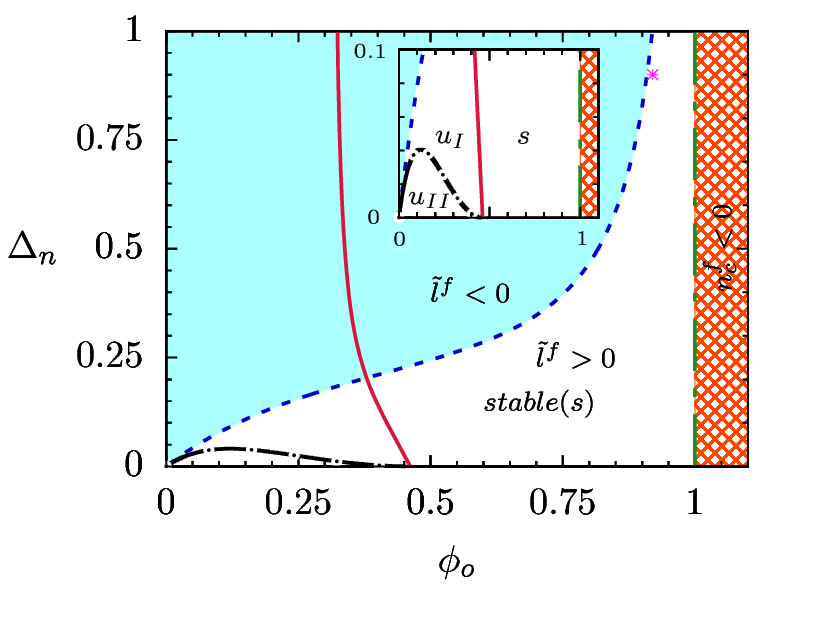}
\caption{Phase diagram of the system in ${\phi}_o-{\Delta}_n$ plane, when there is no interaction between passive cross-linkers and the molecular motors ($\alpha_o = 0$).  Solid (red) line is the linear stability boundary. Dashed (blue) line corresponds to $\tilde{l}_f = 0$.The dotted (green) line is corresponds to $n_{c}^{f} = 0 $. No physical fixed point of $n_c$ exists on the right of green dashed curve. $s$ denotes the region with stable overlap with $n_c^f > 0$ and $l^f > 0$. $u_I$ and $u_{II}$ are the subregions within the linearly unstable region where the fluctuations about the fixed points grows with and without oscillations respectively. Here $\tilde{f_s} = 1.9$, $\gamma = 1$, $\tilde{\Gamma} = 1$.}
\label{fig:1MT}
\end{center}
\end{figure}

\subsection {Non-interacting motor -passive protein-MT system:}
The passive crosslinker entropy is the only contribution to the stability of the complex in this regime. The steady state state solutions, which correspond to the fixed points of Eqs.(4)-(6), read
\begin{eqnarray}
n_{c}^{f} &=& \phi_{o}/g \nonumber\\
n_{b}^{f} &=& \phi_{o}\exp ( g )/\gamma g \nonumber\\
\tilde l^{f} &=& \phi_o \exp ( g )/\gamma g  \Delta_n-2g -4\tilde f_{s} \phi_o/\tilde \Gamma
\end{eqnarray}
where $g = \tilde f_{s} \left ( 1 - \frac{\phi_o}{\tilde \Gamma}\right)$. Since $n_{c}^{f} >0$,  all steady state morphologies must satisfy $\phi_o < \tilde \Gamma$. Analogously, the restriction that  $\tilde l > 0  $ implies  $\Delta_{n} < \Delta_{n}^{c}$, which implicitly depends on  the force $\tilde F_p$. A linear stability analysis around these steady solutions leads to the eigenvalue equation which is of the form,
\begin{eqnarray}
\lambda^{3} + a_{1}\lambda^{2} + a_{2}\lambda + a_3 = 0\nonumber
\end{eqnarray}

\begin{figure}[t]
\begin{center}
\includegraphics[scale = 1.0]{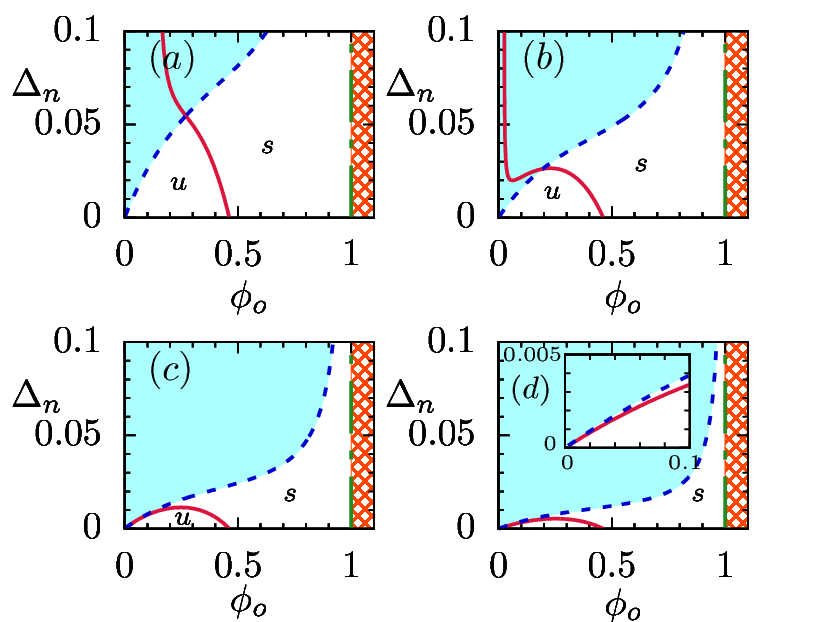}
\caption{Effect of variation of $\gamma$ on stability: $(a)~\gamma=3$, $(b)~\gamma=5$, $(c)~\gamma=10$ and $(d)~\gamma=20$. $u$ denotes the unstable region, while $s$ indicates stable region. Here $\alpha_o = 0$  (No interaction between passive cross-linkers and motors). Other parameter values are, $\tilde{f_s} = 1.9$, $\tilde{\Gamma} = 1$.}
\label{fig:1MT}
\end{center}
\end{figure}

\begin{figure}[t]
\begin{center}
\includegraphics[scale = 1.0]{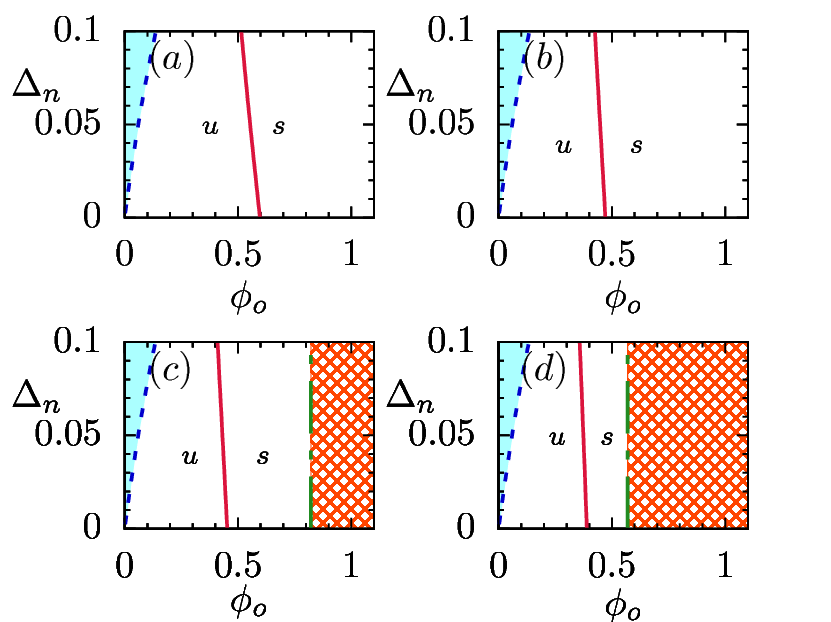}
\caption{Effect of interaction strength $\alpha_o$ on stability: $(a)~\alpha_o=-0.2$ and $(b)~\alpha_o=-0.02$ corresponds to the attractive interaction between crosslinked motors and passive crosslinkers, while  $(c)~\alpha_o=0.02$ and $(d)~\alpha_o=0.2$ corresponds to repulsive interaction between them. $u
$ denotes the unstable region, while $s$ indicates stable region. Here $\tilde{f_s} = 1.9$, $\tilde{\Gamma} = 1$ and $\gamma = 1$.}
\label{fig:1MT}
\end{center}
\end{figure}

\begin{figure}[t]
\begin{center}
\includegraphics[height = 4.5cm, width =8.5cm ]{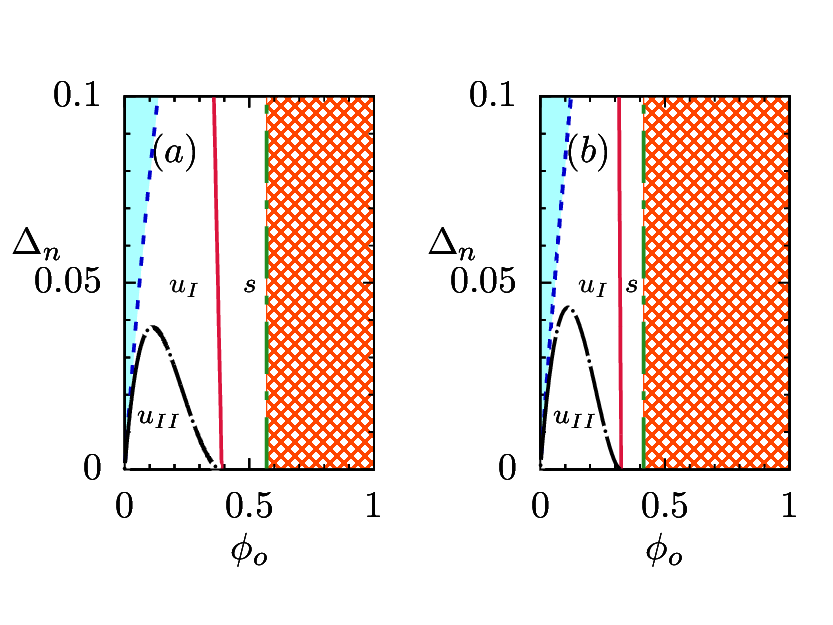}
\caption{Comparison of the phase diagrams for the cases $(a)$ when the passive cross-linkers $(a)$ interact only with crosslinked molecular motors and $(b)$ when the passive crosslinkers interact with both bound and crosslinked motors. $u$ denotes the unstable region, while $s$ denotes region of stable physical overlap. Here $\tilde{f_s} = 1.9$, $\tilde{\Gamma} = 1$ $\gamma = 1$ and $\alpha_o=0.2$.}
\label{fig:1MT}
\end{center}
\end{figure}

\begin{figure}[t]
\begin{center}
\includegraphics[scale = 1.0]{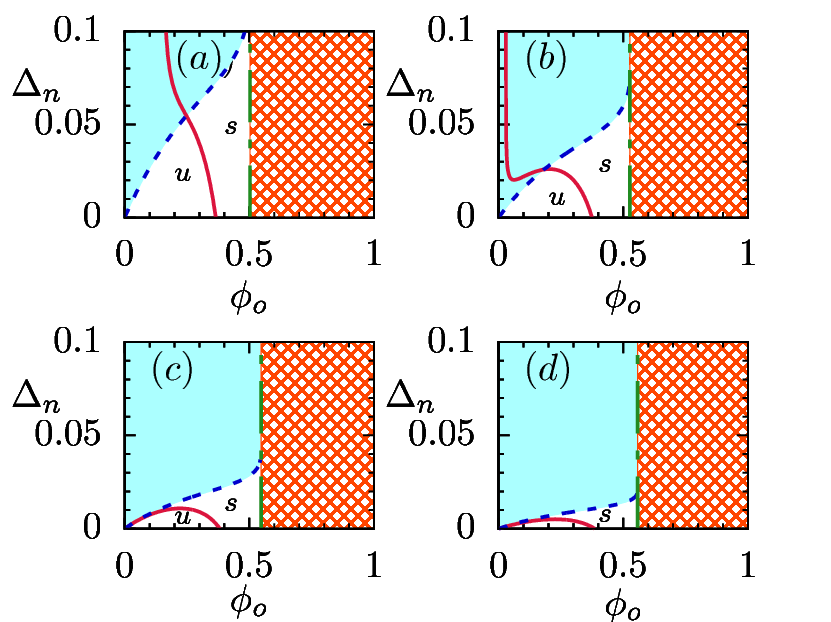}
\caption{Effect of variation of $\gamma$ on stability when passive crosslinkers interact with both crosslinked and bound motors: $(a)~\gamma=3$, $(b)~\gamma=5$, $(c)~\gamma=10$ and $(d)~\gamma=20$ . Here $\tilde{f_s} = 1.9$, $\tilde{\Gamma} = 1$ and $\alpha_o = 0.2$. $u$ denotes the unstable region, while $s$ indicates stable region.}
\label{fig:1MT}
\end{center}
\end{figure}
When one of the roots is real and negative, the other two are complex conjugate, the stability boundary corresponds to the curve $ a_3 = a_2 a_1$ as the real part of the complex conjugate pair changes sign \footnote{The other possible sets of roots for the eigenvalue equation are $(i)$ all roots are real and negative, $(ii)$ one positive real root and a pair of complex conjugate roots and $(iii)$ all roots are real and at least one is positive. While $(i)$ always corresponds to the stable fixed points, both $(ii)$ and $(iii)$ corresponds to  unstable fixed points.}. Accordingly, the  phase boundary separating stable from unstable morphologies of anti-parallel overlapping MTs, reads
\begin{equation}
\Delta_n = \frac{\phi_{o} G}{2g^{2}\gamma}\left( 1 + \frac{1}{\gamma - G} \right)
\label{eqn:pb}
\end{equation}
where $G = (g-1)\exp (g)$  is a function of $\phi_o \equiv l_{e}\phi$. On the  linear stability boundary the real part of the relevant complex eigenvalue vanishes. Hence the pure imaginary eigenvalues lead to a finite  frequency $f_\omega$, which  characterizes the intrinsic oscillations of the MT complex at the threshold of linear stability. The expression for $f_\omega$ reads as, 
\begin{equation}
f_{\omega} =\frac{1}{2\pi}\sqrt{\frac{2g^{2}\Delta_n\gamma}{\phi_o} - G} \nonumber
\end{equation}
As shown in Fig.2, linear stability analysis has allowed us to  span the whole phase diagram of the anti-parallel MT complex as a function of concentration of passive crosslinkers, $\phi_o$, and $\Delta_n$, a measure of motor flux in the overlap region. 
 Overlapping anti-parallel MTs requires  a positive overlap $\tilde{l}^f>0$, 
which restricts the physically meaningful region of phase diagram,  and leads to an upper bound for the strength of the motor flux $\Delta_n^c = \frac{\phi_o e^g \tilde \Gamma}{\gamma g(2g\tilde{\Gamma} + 4\tilde f_s \phi_o)}$ as function of passive crosslinker density.

In Fig. 2, the unshaded region corresponds to  physically meaningful overlapping MTs for which both $(\tilde{l}^f >  0)$ and  $n_c > 0$. This region is  bounded by the dashed blue curve, above which  $(\tilde{l}^f < 0)$, and dash-dotted green line, beyond which  $n_c < 0$.  A positive number of crosslinked motors requires $\phi_{o} < \phi_{o}^{c}$, where  $\phi_{o}^{c} = \tilde \Gamma$. The critical concentration $\phi_{o}^{c}$ depends only on the scaled friction coefficient $\tilde \Gamma$ and  does not depend on the (un)binding motor rates.The red curve separates the linearly stable regime of the anti-parallel MT arrangement with the unstable regime. The linearly unstable region can be further classified into two sub-regions based on how the small deviations about the fixed points grow with time. While in one region the instability grows with a characteristic frequency, in the other the instabilities grow monotonously in time.

Fig.3 depicts the effect of $\gamma$ on the stability of the overlapping MTs. Increasing $\gamma$ diminishes the region of stable physical overlap. It is worthwhile to point out that although the region of stable overlap diminishes, beyond a threshold value of $\gamma$, even for arbitrary small concentration of passive proteins, $\phi_o$, there always exists a range of $\Delta_n$ for which the overlap configuration of the MTs can be stabilized, as illustrated in the insets of  Fig.3c and Fig.3d.

\subsection {Interacting motor -passive protein-MT system}: From Eqs.(7-9) we can identify the two distinct sets of fixed points  that quantify the steady states when there is mutual interaction only between the passive crosslinkers and crosslinked motor proteins, which read as, 

\begin{eqnarray}
n_{c}^{f} &=& 1/\omega_{\pm}= \left\{- \eta \pm \sqrt{{\eta^{2} - 4\alpha_{o}} \phi_{o}^{2} \tilde f_{s} \tilde \Gamma }\right\}/2\alpha_{o}\phi_{o} \tilde f_{s} \nonumber \\
n_{b}^{f} &=& \exp (\phi_{o} \omega_{\pm} + \alpha_o \phi_o)/\gamma \omega_{\pm} \nonumber\\ 
\tilde l^{f} &=& \exp(\phi_{o}\omega_{\pm} + \alpha_o \phi_o) /\gamma \Delta_{n} \omega_{\pm}- 2 \phi_{o} \omega_{\pm} - 2 \alpha_o \phi_o \nonumber\\
&~& -~4 \tilde f_{s} \phi_o/\tilde \Gamma - 4 \tilde f_{s} \alpha_{o} \phi_o/\omega_{\pm} \tilde \Gamma \nonumber
\end{eqnarray}
where $\eta~=~\tilde \Gamma (\alpha_o\phi_o - g )$

For attractive interactions between passive crosslinkers and motors ( $\alpha_o < 0$), The solution of fixed point corresponding to $\omega_{+}$ cannot be sustained  since $\omega_{+} $ is always negative, leading to   $n_c^f<0$. 
For repulsive interactions ($\alpha_o > 0$), the solution of fixed point corresponding to $ \omega_{+}$ leads  to an unstable fixed point. \footnote {When $\alpha_{o} > 0$, the set of fixed points corresponding to $ \omega_{+}$ has always at least one real positive root for the eigenvalue equation for the fluctuation spectrum which corresponds to linearly unstable fixed point.} Thus only the solution corresponding to $\omega_{-}$ leads to  physically relevant steady states. The corresponding expression for stability boundary is,
\begin{equation}
\Delta_n = \frac{\Omega \left( 1 - \frac{1}{\Omega - \gamma}\right)}{2\gamma \left[ \omega_{-}^{2} \phi_{o} + \frac{\alpha_{o} \phi_{o} \tilde f_s}{\tilde \Gamma}\left(\frac{1}{\Omega - \gamma} - 2 \right) \right]}
\end{equation}

where, $\Omega = (\phi_o \omega_{-} - 1 ) \exp (\phi_o \omega_{-} + \alpha_o\phi_o)$. In this regime also intrinsic, spontaneous, oscillations develop at the boundary of stability of magnitude
\begin{equation}
f_{\omega} =\frac{1}{2\pi}\sqrt{2\phi_o\omega_-^2\Delta_n\gamma - \frac{4\alpha_o\phi_o\Delta_n\tilde{f}_s\gamma}{\tilde{\Gamma}}-\Omega} 
\end{equation}
 
Fig.4  displays the phase diagram in $\phi_{o}- \Delta_n$ plane for different interaction strengths, $\alpha_o$,  when  passive proteins only  interact with crosslinked motors. Both for  attractive, $\alpha_o < 0$, and repulsive, $\alpha_o >0$, interactions , increasing the interaction, $\alpha_o $,  reduces the region of stable physical overlap of MTs. Significantly, when $\alpha_o < 0$, there is no upper bound of passive crosslinkers, $\phi_o$ (Fig.4a and Fig.4b), beyond which a stable MT physical overlap  can be obtained. This is in contrast to the situation for which there is no interaction ($\alpha_o = 0$) as illustrated in Fig.1 or when there is attractive interaction between the crosslinked motors and crosslinking passive proteins as shown in Fig. 4c and Fig.4d, for which there is an upper bound of $\phi_o$, beyond which physical overlap of MTs cannot be obtained.   

Fig.5a and Fig.5b shows the comparison of  the phase diagrams for the case when passive proteins interact  {\it only} with crosslinked motors with the case for which the passive crosslinkers interact with both bound and crosslinked motors. For the latter case, there is substantial reduction of the region in the phase diagram for which  physical stable overlapping MTs can be obtained.  Finally, Fig.6 shows that increasing $\gamma$ decreases the region where overlapping MTs are stable when passive proteins interact with both the bound and the crosslinked motors. 

\subsection {Experimental relevance:} We can estimate the typical overlap length of MTs and motor numbers using experimental data known for biological systems. For single motor velocity, $v_o = 1~\mu$m$s^{-1}$,\cite{lipo-bd1,lipo-uni} and the bare unbinding rate $k_u^o = 1~s^{-1}$ \cite{lipo-bd1,lipo-uni}, the typical processivity length $l_p = 1~\mu$m. Using the value of thermal energy, $k_B$T$=4.2$ pN-nm~\cite{abhishek2016}, and length scale associated with unbinding process of motors, $b = 1.3$ nm~\cite{schnitzer2000}, an estimate for the additional reference length scale, $l_e = 2.6$ nm is obtained. Experimental measurements for the stall force for kinesin, $f_s= 6$ pN~\cite{lipo-bd1,lipo-uni}, in turn implies $\tilde{f}_s \sim 1.9$. For an experimental estimate of the  binding rate, $k_b = 5~s^{-1}$\cite{lipo-bd1,lipo-uni}, we get $\gamma \sim 5$. Indeed under different physiological conditions, $\gamma$ may vary one order of magnitude~\cite{abhishek2016}. Furthermore, we consider that  passive crosslinkers  bind with a characteristic energy, $\epsilon_o=2k_B$T. Quantitative estimates of the friction coefficient   of passive crosslinkers at low concentrations \cite {lansky1} implies that  the scaled friction coefficient $\tilde{\Gamma}\sim 1$. Assuming a linear density  of  passive crosslinker, $\phi = 0.35$ nm$^{-1}$, and choosing the bath  motor concentration  such that ${\rho}_{3D}k_b^{3D} = 1.7~{\mu}$m$^{-1}$ s$^{-1}$ (corresponding to $\phi_o = 0.9$, $\Delta_n = 0.9$, the point is marked with  `$\ast$' in Fig.2) we obtain an estimate of the stable overlap length $\sim 0.45~\mu$m with bound and crosslinked motor numbers being  $n_b  \sim 7$, and $n_c \sim 6$, respectively. These estimates fall well within the scale of cellular processes and thus suggests that the interplay between the active forces due to motors and the passive crosslinkers is a biologically relevant mechanism in determining the stability of overlapping MT filaments.
 
 \section{Summary and Discussion} In summary we have shown that  the presence of passive crosslinkers qualitatively impacts the mechanical properties of  arrangements of overlapping   biofilaments that are subject to active sliding forces due to molecular motors. These forces arise due to the  interaction of passive crosslinkers with overlapping filaments in the presence of a passive crosslinker bath. This is a distinct mechanism from other passive force generating mechanism such as entropic forces due to confinement or forces due to depletant molecules. We have shown that for a pair of anti-parallel MTs, the interplay of sliding forces due to such passive proteins crosslinkers with active sliding forces due to molecular motors can stabilize  the overlapping filament pair. Further we have identified  that  this passive mechanism  is necessary to ensure   the emergence of  stable overlaps. The effect of attractive mutual interaction between  passive crosslinkers and  active motors leads to an enhanced range of passive crosslinker density for which stable overlapping MT pair can be obtained while a repulsive interaction between the motors and passive crosslinkers tend to diminish the domain of stable overlaps.

{\bf Acknowledgment}\\
IP acknowledges MINECO and DURSI for financial
support under projects FIS2015-67837- P and 2017SGR844,
respectively. SM acknowledges SERB project EMR/2017/001335 for financial support.


\begin{thebibliography}{}

\bibitem{cell} B. Alberts at. al , {\sl Molecular Biology of the cell} ( Garland Science, New York, 2007, 6th ed)

\bibitem{howard} J. Howard, {\sl Mechanics of motor proteins and the cytoskeleton} (Sinauer Associates, Sunderland, 2001)

\bibitem{nedelec} F. Nedelec, T. Surrey, A.C. Maggs, and S. Leibler, Nature {\bf 389}, 305 (1997)

\bibitem{madan} Y. Hatwalne, S. Ramaswamy, M. Rao, and R. A. Simha, Phys. Rev. Lett. {\bf 92}, 118101 (2004)

\bibitem{menon} S. Sankararaman, G. I. Menon, and P. B. Sunil Kumar,  Phys. Rev. E {\bf 70}, 031905 (2004)

\bibitem{raja1} N. P. Ferenz, R. Paul, C. Fagerstrom, A. Mogilner, and P. Wadsworth, Curr. Biol. {\bf 19} 1833 (2009)

\bibitem{sm1} S. Muhuri, I. Pagonabarraga, and J. Casademunt, EPL {\bf 98}, 68005(2012)

\bibitem{sm2} P. Malgaretti and S. Muhuri, EPL {\bf 115}, 28001 (2016)

\bibitem{debasoft} S. Ghosh, V. N. S. Pradeep, S. Muhuri, I. Pagonabarraga, and D. Chaudhuri, Soft Matter {\bf 13}, 7129 (2017)

\bibitem{karsten1} S. W. Grill, K. Kruse, and F. Julicher, Phys. Rev. Lett. {\bf 94}, 108104 (2005)

\bibitem{kinesin1} X. Su, H. Arellano-Santoyo, D. Portman, J. Gailard, M. Vantard, M. Thery, and D. Pellman, Nat. Cell. Biol. {\bf 15}, 948 (2013)

\bibitem{kinesin2} Y. Fukuda, A. Luchniak, E. R. Murphy, and M. L. Gupta, Curr. Biol. {\bf 24}, 1826 (2014)

\bibitem{lansky1} Z. Lansky, M. Braun, A. Ludecke, M. Schlierf, P. R. ten Wolde, M. E. Janson, and S. Diez, Cell {\bf 160}, 1159 (2015)

\bibitem{dogic_review} M. Braun, Z. Lansky, F. Hilitski, Z. Dogic, S. Diez, Bioessays {\bf 38}, 474 (2016)

\bibitem{lk1} R. Subramanian, E. M. Wilson-Kubalek, C. P. Arthur, M. J. Bick, E. A. Campbell, S. A. Darst, R. A. Milligan, and T. M. Kapoor, Cell {\bf 142}, 433 (2010)

\bibitem{lk2} P. Bieling, I. A. Telley, and T. Surrey, Cell {\bf 142}, 420 (2010)

\bibitem{lk3} H. S. Kuan and M. D. Betterton, Biophys. J. {\bf 110} 2034 ( 2016)

\bibitem{lansky2} M. Braun, Z. Lansky, F. Hilitski, Z. Dogic, and S. Diez, Bioessays {\bf 39}, 474 (2016)

\bibitem{lansky3} Z. Lansky, M. Braub, A. L\"udecke, M. Schlierf, P.R. ten Wolde, M.E. Janson, S. Diez, Cell {\bf 160}, 1159 (2015) (and comment by D. J. Odde, Cell {\bf 160} 1041 (2015)).

\bibitem{dogic_prl} F. Hilitski, A.R. Ward, L. Cajamarca, M.F. Hagan, G.M. Grason, Z. Dogic, Phys. Rev. Lett. {\bf 114}, 138102 (2015)

\bibitem{kruse_passive} D. Johann, D. Goswami, K. Kruse, Phys. Rev. E {\bf 93}, 062415 (2016)

\bibitem{karsten2} D. Johann, D. Goswami, and K. Kruse, Phys. Rev. Lett. {\bf 115}, 118103 (2015)

\bibitem{ignaref} O.Campas, J. Casademunt, and I. Pagonabarraga, EPL {\bf 81}, 48003 (2008) 

\bibitem{lipo-uni} S. Klumpp and R. Lipowsky, Proc. Natl. Acad. Sci. {\bf 102}, 284 (2005)

\bibitem{lipo-bd1} M. J. I. Muller, S. Klumpp and R. Lipowsky, Proc. Natl. Acad. Sci. {\bf 105}, 4609 (2008)

\bibitem{schnitzer2000} Schnitzer M. J. et al., Nat. Cell Biol., {\bf 2},  718 (2000)

\bibitem{abhishek2016} A. Chaudhuri and D. Chaudhuri, Soft Matter {\bf 12},  2157  (2016)

\end{thebibliography}
\end{document}